\documentclass[prb,superscriptaddress,showpacs,aps]{revtex4}% Physical Review B
\usepackage{color}
\usepackage{graphicx}% Include figure files
\usepackage{dcolumn}% Align table columns on decimal point
\usepackage{bm}% bold math
\usepackage{amsmath, amsthm, amssymb, amsfonts}

\begin{document}

\preprint{APS/123-QED}

\title{Stability properties of a thin relativistic beam propagation in a magnetized plasma}
\author{Du\v san Jovanovi\'c}
\email{dusan.jovanovic@ipb.ac.rs} \affiliation{Institute of Physics, University of Belgrade, Pregrevica 118, 11080 Belgrade (Zemun), Serbia}
\author{Renato Fedele}
\email{renato.fedele@na.infn.it} \affiliation{Dipartimento di Fisica, Universit\`{a} di Napoli "Federico II" Complesso Universitario
M.S. Angelo,
%via Cintia, I-80126
Napoli, Italy}
\affiliation{INFN Sezione di Napoli, Complesso Universitario di M.S. Angelo, Napoli, Italy}
\author{Milivoj Beli\'c}
\email{milivoj.belic@qatar.tamu.edu} \affiliation{Texas A\&M University at Qatar, P.O. Box 23874 Doha, Qatar}
\author{Sergio De Nicola}
\email{sergio.denicola@spin.cnr.it} \affiliation{SPIN-CNR, Complesso Universitario di M.S. Angelo, Napoli, Italy}
\affiliation{Dipartimento di Fisica, Universit\`{a} di Napoli "Federico II" Complesso Universitario
M.S. Angelo,
%via Cintia, I-80126
Napoli, Italy}

\date{\today}

\begin{abstract}
A self-consistent nonlinear hydrodynamic theory is presented of the propagation of a long and thin relativistic electron beam through a plasma that is relatively strongly magnetized, $|\Omega_e|\sim\omega_{pe}$ and whose density is much bigger than that of the beam. In the regime when the parallel phase velocity in the comoving frame is much smaller than the thermal speed and the beam electrons are thermalized, a stationary solution for the beam is found when the electron motion in the transverse direction is negligibile and the transverse localization comes from the nonlinearity of its 3-D adiabatic expansion. Conversely, when the parallel phase velocity is sufficiently large to prevent the heat convection along the magnetic field, a helicoidally shaped stationary beam is found whose transverse profile is determined from a nonlinear dispersion relation and depends on the transverse size of the beam and its pitch angle.
\end{abstract}

% PACS, the Physics and Astronomy Classification Scheme.
%\keywords{Suggested keywords}%Use showkeys class option if keyword display desired

\maketitle

\section{Introduction}\label{Introduction}

The propagation of thin relativistic electron or positron beams along a strong magnetic field immersed in a plasma has attracted a lot of interest in the last decades. Besides the fundamental theoretical importance of the processes involved in such interaction, the interest has been stimulated also by the possibility of their practical applications. The experimental configurations of such type have been proposed for the generation of intense X-rays in free electron lasers (FEL), see e.g. \cite{2012ApPhL.100i1110H} and references therein, and also for the new concept of plasma based particle accelerators that include the plasma wake field accelerator (PWFA) \cite{1985PhRvL..54..693C} and the inverse FEL schemes \cite{2014NatCo...5E4928D,2015NatSR...515499H}. Moreover, the radiation from moving, localized charge distributions plays an important role in the generation and trapping of whistler waves in the radiation belts of the Earth's magnetosphere, for the auroral electron beams in the ionosphere, for the current and temperature filaments in tokamaks, etc.

Plasma-based particle accelerators are expected to be capable of accelerating charged particles beyond 10 TeV. The PWFA concept consists in the injection of a relativistic electron or positron beam into the plasma, creating a vacancy in the electron population of the background plasma, while the much heavier plasma ions remain in place. In the wake of the driver beam, a very large amplitude electrostatic wave is created that propagates with the same velocity as the beam. When the latter is properly bunched and the beam density is of the order of $10^{17}-10^{18}\, {\rm cm}^{-3}$, an accelerating electric with the intensity up to $1$ GV/cm can be created, yielding the energy of the accelerated particles to be $3-4$ times bigger that that of the driving beam \cite{1985PhRvL..54..693C}.
The dynamics of the driving beam of a PWFA device, coupled with the electric field created by it inside the plasma in the presence of a magnetic field, was studied in our earlier papers \cite{2014EPJD...68..210F,2014EPJD...68..271F} using the Vlasov's kinetic theory and a macroscopic (i.e. an envelope) description for the beam itself, based on the virial theorem. This has been done in the strongly nonlocal case (when the beam spot-size is much smaller than the plasma wavelength), in the overdense regime (when the beam density is much smaller than that of the plasma), and for a sufficiently long beam, adiabatically shielded by the plasma. It was demonstrated that in the longitudinal direction a driving beam of finite extent can undergo a two-stream instability in the longitudinal direction, while in the transverse direction it can be subjected to self-focusing/defocusing, filamentation, or self-pinching. The feedback from the transverse plasma wake field on the driving beam and the thermal spreading of the latter lead to the nonlinear self modulation of the beam, i.e. to the sausage-shaped betatron-like oscillations of the beam envelope whose period depends on the external magnetic field. On longer time scales and under suitable conditions, these can be either stable or unstable (so-called self modulation instability). However, the macroscopic beam description of Refs. \cite{2014EPJD...68..210F,2014EPJD...68..271F} based on the virial theorem, assumed a cylindrically symmetric structure and did not account for any transverse dynamics of the beam's centroid, including possible displacements in the transverse direction. When the beam transverse displacements are one dimensional (1-D), i.e. when the beam motion occurs in a plane parallel to the direction of propagation, the hose instability is excited with a growth rate that is comparable to that of the self-modulation \cite{2012PhRvE..86b6402S}. Earlier models suggested that the hose instability would significantly alter the structure of the plasma wake field and limit the applicability of plasma-wake field accelerators, but recently it has been shown \cite{2017PhRvL.118q4801M} that the inherent drive-beam energy loss, along with an initial beam-energy spread, detunes the betatron oscillations of beam electrons and thus quenches the instability, stabilizing the drive beam over long propagation distances.

Besides the 1-D beam excursions that lead to the hose instability, it is necessary to study also 2-D transverse displacements such as those existing in a helicoidal geometry, because of the unique advantages of helicoidal electron bunches for the use in FELs and in inverse FEL accelerator schemes, as well as in direct laser acceleration (DLA) that may increase the efficiency of LWA.
In a DLA scheme, the electrons  inside the ion channel are trapped by the combination of the longitudinal external magnetic field and the strong self-generated quasi-static magnetic and %electric, i.e.
ponderomotive fields. In the transverse direction, trapped electrons undergo betatron-like oscillations in the self-generated fields while moving on (almost) circular trajectories with fixed radii. The orbits of trapped electrons in phase space are closed, the variation of their ponderomotive phases is very slow, and consequently they stay longer in the acceleration phase and get accelerated more efficiently. The generation of a helical electron bunch via direct laser acceleration, using a circularly polarized laser pulse propagating in a plasma with a near-critical density, has been demonstrated in 3-D self-consistent Particle-in-Cell (PIC) simulations \cite{2015NatSR...515499H}. Likewise, an inverse FEL scheme concocted as a combination of PWFA and LWA \cite{2014NatCo...5E4928D}, that utilized the interaction between a helical electron beam (driven through a $0.5\,{\rm m}$ long tapered helical undulator) and a laser beam with a modest intensity $\sim 10^{13}\,{\rm W/cm}^2$, produced a wake with an accelerating gradient $>0.1\, {\rm GV/m}$ delivering a $450\, {\rm MeV}$ energy gain with excellent output beam quality. Helicoidal electron bunches were created also using a circularly polarized laser beam and a helical undulator \cite{2012ApPhL.100i1110H}. The emergence of microbunching in this experiment was documented by the observed coherent transition radiation corresponding to a dominantly helical electron beam density distribution. Such bunches can be used as tunable sources of coherent light with orbital angular momentum in high-gain free-electron lasers.

The radiation from helicoidal bunches entering plasma from a vacuum, or propagating in a plasma, were studied by \cite{2009JAP...105i3101H,1996PhPl....3.4717R}, neglecting the feedback from the plasma to the driver bunch, assuming that the plasma is relatively weakly magnetized $|\Omega_e|\ll\omega_{pe}$. Frequencies above $|\Omega_e|$ were considered, for which three regimes were identified. In the \textit{evanescent regime} $|\Omega_e|<\omega <\omega_{pe}$, plasma modes are evanescent in the perpendicular direction and localized near the beam. As a consequence of the helical geometry, spatially localized magnetic fluctuations are generated at the harmonics of the electron-cyclotron frequency whose width is determined by the pitch angle of the beam. In the \textit{upper hybrid regime} $\omega_{pe}<\omega <\omega_{uh}$, the beam excites propagating upper hybrid waves, and in the \textit{vacuum regime} $\omega> \omega_{uh}$ the beam excites propagating electromagnetic modes whose frequencies are near the harmonics of $\Omega_e$.

In the present paper we present a self-consistent nonlinear hydrodynamic theory of the propagation of a long and thin relativistic electron beam through a plasma. The plasma is overdense (i.e. its density is much bigger than that of the beam) and relatively strongly magnetized, $|\Omega_e|\sim\omega_{pe}$. Two different regimes of the beam evolution are considered with respect to the rate of parallel heat convection. Firstly, if the parallel phase velocity in the comoving frame is much smaller than the thermal speed, the beam electrons are thermalized and a stationary solution for the beam profile is found in the regime when the electron motion in the transverse direction is negligibile. For such a beam, the localization in the transverse direction is governed by the nonlinearity associated with its 3-D adiabatic expansion while the self-modulation and the related betatron emission are weak. Conversely, when the parallel phase velocity is sufficiently large to prevent the heat convection along the magnetic field, the beam can have an arbitrary cylindrically symmetric transverse profile that remains stable on the timescale that is short compared to the period of betatron oscillations. However, if  the beam is launched at a small pitch angle relative to the magnetic field so that it attains a helicoidal shape, it remains stable only if its transverse profile satisfies a nonlinear dispersion relation determined by the transverse size of the beam and its pitch angle.

\section{The description of the plasma}\label{plasma_descr}

The system that we study in the present paper consists of a quiescent plasma with the homogeneous unperturbed density is $n_{p 0}$, immersed in a homogeneous magnetic field $\vec{e}_z B_0$ and pierced by a relativistic beam of electrons with the density $n_b$. In the rest of the paper, the plasma and the beam parameters are denoted by the subscripts $p$ and $b$, respectively. The beam propagates with a relativistic velocity $\vec{e}_z u$, with $c-u\ll c$, and it is relatively long, $L_{b\Vert}\gg L_{b\bot}$. Here $c$ is the speed of light and $L_{b\Vert}$ and $L_{b\bot}$ are the characteristic lengths of inhomogeneity of the beam in the directions parallel and perpendicular to its direction of propagation, respectively. The electromagnetic field in such a system is described by the standard wave equation
\begin{equation}\label{wave_equation}
\nabla\left(\nabla\cdot\vec E\right)-\nabla^2\vec E + \frac{1}{c^2}\frac{\partial^2\vec E}{\partial t^2} = -\frac{1}{c^2\epsilon_0}\frac{\partial}{\partial t}\left(\vec{j}_p + \vec{j}_b\right),
\end{equation}
where the electric current is the sum of the contributions of the plasma- and beam electrons, viz. $\vec j_p = -e n_p \vec{v}_p$ and $\vec j_b = -e n_b \vec{v}_b$; here $n_l$ and $\vec{v}_l$ ($l=p,b$) are the fluid density and the fluid velocity of the species $l$. We take that both the velocity of the plasma electrons and the deviation of the velocity of the beam electrons from their unperturbed velocity $\vec{e}_z u$ are nonrelativistic, viz. $|\vec{v}_p|\sim |\vec{v}_b - \vec{e}_z u|\ll c$. The density of the beam is assumed to be much smaller than that of the background plasma, $n_b/n_p \sim v_p/u \ll 1$, so that the plasma and beam currents are of the same order. For such overdense plasma and a not too strong electromagnetic field, in which the potential energy of an electron in the electrostatic potential $\phi$ and in the $z-$ component of the vector potential $A_z$ is much smaller than the electron rest energy, viz. $\phi\sim u A_z\ll m_0 c^2/e$, $m_0$ being the electron rest mass. Introducing the small parameter $\varepsilon\ll 1$ as
\begin{equation}\label{scaling}
\varepsilon\sim \frac{c-u}{c}\sim \frac{n_b}{n_{p0}}\sim\frac{e\left(\phi-u A_z\right)}{m_0c^2}\ll 1,
\end{equation}
we can now rewrite the $i$-th component of the Fourier-transform of the wave equation (\ref{wave_equation}), with the accuracy to the leading order in $\varepsilon$, in the following familiar form
\begin{equation}\label{Fourier_wave_eq}
\left(k^2 \,\delta_{i,j} -k_i k_j -\frac{\omega^2}{c^2}\; \epsilon_{i,j}\right)E_j = -\frac{i\omega}{c^2\epsilon_0} \; e n_b\, u \;\delta_{i,z}
\end{equation}
where $\delta_{i,j}$ is the Kroenecker delta and $\epsilon_{i,j}$ is the $i,j$ component of the dielectric tensor $\bm\epsilon$, viz.
\begin{equation}\label{tensor_epsilon}
{\bm \epsilon} =
\left\|
\begin{array}{ccc}
\epsilon_1       & i \, \epsilon_2 & 0 \\
-i \, \epsilon_2 & \epsilon_1      & 0 \\
0                & 0               & \epsilon_3
\end{array}
\right\|,
\quad  \quad
\epsilon_1 = 1 - \frac{\omega_{pp}^2}{\Omega_{p}^2-\omega^2}, \quad \epsilon_2 = \frac{\Omega_p \, \omega_{pp}^2}{\omega\left(\Omega_{p}^2-\omega^2\right)}, \quad \epsilon_3 = 1 - \frac{\omega_{pp}^2}{\omega^2},
\end{equation}
where $\omega_{pp} = \sqrt{n_{p0} e^2/m_0\epsilon_0}$ and $\Omega_p = -e B_0/m_0$ are the plasma frequency and the gyrofrequency of the plasma electrons, respectively. The dielectric tensor (\ref{tensor_epsilon}) is pertinent to high frequency perturbations whose parallel phase velocity is comparable to the speed of light,
\begin{equation}\label{frequency_scaling}
\omega\sim c  k_z\ll c k_\bot\sim\omega_{pp}\sim\Omega_p,
\end{equation}
for which both the ion dynamics and the electron thermal effects can be neglected within the adopted accuracy. Here $\bot$ stands for the vector components perpendicular to the direction of propagation of the beam.

Solving Eqs. (\ref{Fourier_wave_eq}) for the parallel electric field, which is conveniently expressed as $E_z=-i k_z[\phi-(\omega/k_z)A_z]$, where $\phi$ and $A_z$ are the electrostatic potential and the $z$-component of the vector potential, respectively, we have
\begin{equation}\label{res_lin_eq}
\phi-\left(\omega/k_z\right)A_z = \frac{R}{D}\frac{n_b}{n_{p0}},
\end{equation}
where $D$ is the determinant of the dispersion matrix ${\bm D}$, defined as $D_{i,j} = k^2 \,\delta_{i,j} -k_i k_j -(\omega^2/c^2)\; \epsilon_{i,j}$, viz.
\begin{eqnarray}
\nonumber
D &=& c^4 k_\bot^4 \Omega_p^2 + (\omega_{pp}^2 - \omega^2)(\omega_{pp}^2 + c^2 k_\bot^2)^2 +
\\
\label{lin_disp_rel}
&&
\frac{\omega^2 - c^2 k_z^2}{\omega^2} \left\{\left(\omega^2 - \omega_{pp}^2\right) \left[\omega^2 \left(c^2 k_\bot^2 + 2 \omega_{pp}^2\right) - \left(\omega^2 - c^2 k_z^2 -
c^2 k_\bot^2\right) \left(\omega^2 - \Omega_p^2\right)\right] - \omega^2 \Omega_p^2 c^2 k_\bot^2\right\}
\\
\nonumber
R &=& \omega_{pp}^2 \;\frac{m_0 \omega \, u}{e\,k_z}%\frac{n_b}{n_{p0}}
\left\{\left(\omega^2 - \omega_{pp}^2\right)^2 + c^2 k_\bot^2 \omega_{pp}^2 + \right.
\\
&&
\left. \frac{\omega^2 - c^2 k_z^2}{\omega^2}\; c^2 k_\bot^2\left(\Omega_p^2 - \omega^2\right)-
2 c^2 k_z^2\left[\omega^2 - \Omega_p^2 - \omega_{pp}^2 - \frac{c^2 k_z^2}{2 \omega^2} \left(\omega^2 - \Omega_p^2\right)\right] - \omega^2\Omega_p^2\right\}
\label{QQQ}
\end{eqnarray}
The dispersion relation $D=0$ describes linear electromagnetic perturbations whose frequency $\omega$ is sufficiently high so that the ions can be regarded as immobile, i.e. for $\omega\gg{\rm max}(\Omega_i, \omega_{pi})$, where $\Omega_i$ and $\omega_{pi}$ are the gyrofrequency and the plasma frequency of the plasma ions, respectively. When its mode frequency is also small compared to the gyrofrequency of the plasma electrons, $\omega<|\Omega_p|$, see Eq. (\ref{frequency_scaling}), the electromagnetic perturbation is usually referred to as the \textit{obliquely propagating whistler}. For an almost perpendicular propagation envisaged in Eq. (\ref{frequency_scaling}), with the accuracy to the leading order in $k_z/k_\bot\sim\omega/ck_\bot$, the dispersion relation $D=0$ is readily solved as
\begin{equation}
\label{oblique}
\frac{c^2k_\bot^2}{\omega_{pp}^2} =
\frac{\Omega_p^2 (c^2 k_z^2-\omega^2)-2\omega^2\omega_{pp} \pm \Omega_p \sqrt{\Omega_p^2 \left(c^2 k_z^2 - \omega^2\right)^2 - 4\, c^2 k_z^2 \, \omega^2\omega_{pp}^2}}{2 \omega^2 \left(\Omega_p^2 + \omega_{pp}^2\right)}.
\end{equation}
The  wavenumber $k_\bot$ of such oblique whistler is a real quantity if the parallel phase velocity $\omega/k_z$ is sufficiently small
\begin{equation}
\label{propagates}
\frac{\omega^2}{c^2 k_z^2}< \frac{\sqrt{\Omega_p^2 + \omega_{pp}^2} - \omega_{pp}}{
\sqrt{\Omega_p^2 + \omega_{pp}^2} + \omega_{pp}}<1,
\end{equation}
and otherwise the wave is evanescent. Obviously, for an efficient wake field scheme we need a plasma response that is in resonance with the relativistic beam, i.e. which satisfies $\omega = k_z u + {\cal O}(\varepsilon)\approx k_z c$. From Eq. (\ref{propagates}) we see that such response is also well localized in the transverse direction. Performing the inverse Fourier transform, i.e. making  the substitutions $\omega\to i\;\partial/\partial t$ and $k_j\to -i\;\partial/\partial x_j$ and for a sufficiently long beam, $\partial/\partial z\ll\nabla_\bot$, neglecting the terms of the order $\varepsilon$ and smaller, we obtain the following relation between the potentials $\phi$ and $A_z$ and the beam density $n_b$
\begin{equation}\label{plasma_equation}
\left[\left(1-\frac{c^2\nabla_\bot^2}{\omega_{pe}^2}\right)^2+\frac{\Omega_p^2\; c^4\nabla_\bot^4}{\omega_{pe}^6}\right]\frac{e \varphi}{m_0c^2} = \left(1-\frac{c^2\nabla_\bot^2}{\omega_{pe}^2}\right)\frac{n_b}{n_{p0}},
\quad {\rm where}\quad
\varphi = \phi-u A_z.
\end{equation}
As it will be shown below, the beam density $n_b$ varies on the timescale that is slow compared to $u\, \partial/\partial z$ and, within the adopted accuracy, our Eq. (\ref{plasma_equation}) describes the instantaneous plasma response to such slow variation.

\section{The description of the beam}\label{beam_descr}

The dynamics of the beam is conveniently described by the collisionless hydrodynamic equations written in the reference frame comoving with the beam, which we carry out by introducing the following notations
\begin{equation}\label{shifted_variables}
\vec{r}^{\;\prime} = \vec{r} - \vec{e}_z u t,
\quad
\vec{v}^{\;\prime} = \vec{v}_b - \vec{e}_z u,
\quad
\vec{E}^{\,\prime} = \vec{E} + \vec{e}_z u \times\vec{B},
\quad
\partial/\partial t'= \partial/\partial t + u \; \partial/\partial z.
\end{equation}
Specifically, $\partial/\partial t'$ corresponds to the slow variation in the comoving frame. The appropriate scaling of such slow variation is elaborated in Eq. (\ref{drift_scaling}). The velocity $\vec{v}^{\;\prime}$ is assumed to be nonrelativistic, satisfying $|\vec{v}^{\;\prime}|\ll c-u\ll c$, and consequently the momentum of the beam electrons can be approximated as
\begin{equation}\label{beam_momentum}
m\vec{v}_b \equiv \frac{m_0\left(\vec{v}^{\,\prime}+\vec{e}_z u\right)}{\sqrt{1 - \left(\vec{v}^{\,\prime}+\vec{e}_z u\right)^2 \!\! /c^2}} \approx
%\frac{m_0}{\sqrt{1-u^2/c^2}}\left[\vec{v}_\bot^{\;\prime} + \vec{e}_z \left(u + \frac{v_z'}{1-u^2/c^2}\right)\right] =
\frac{m_0}{\sqrt{1-u^2/c^2}}\left[\vec{v}^{\;\prime} + \vec{e}_z \left(u + \frac{u^2 v_z'}{c^2-u^2}\right)\right],
\end{equation}
which permits us to write the equations of continuity and momentum as follows
\begin{eqnarray}
&&
\frac{\partial n_b}{\partial t'}+\nabla'\cdot\left(n_b \vec{v}^{\;\prime}\right) = 0,
\label{beam_cont}
\\
&&
\left(\frac{\partial}{\partial t'}+\vec{v}^{\;\prime}\cdot\nabla'\right) \vec{v}^{\;\prime\prime}
= -\frac{e}{m_b} \left(\vec{E}^{\,\prime} + \vec{v}^{\;\prime} \times\vec{B}\right) - \frac{1}{m_b n_b}\; \nabla'\cdot\left({\bf p}_b + {\bm\pi}_b\right),
\quad{\rm where}\quad \vec{v}^{\;\prime\prime} = \vec{v}^{\;\prime} + \left(\gamma_u^2-1\right)\vec{e}_z v_z',
\hspace{1cm}
\label{beam_mom}
\end{eqnarray}
and for more details about the foundation of the relativistic hydrodynamic plasma description see e.g. Refs. \cite{2003PhRvL..90c5001M,2014NPGeo..21..217M}.
For later reference, we write also the parallel momentum equation, obtained by the scalar product of Eq. (\ref{beam_mom}) with a unit vector parallel to the magnetic field, $\vec{e}_\Vert$, where $\vec{e}_\Vert = \vec{B}/B$ and $B$ is the intensity of the magnetic field, $B = |\vec{B}|$
\begin{equation}\label{parallel_momentum_eq}
\left(\frac{\partial}{\partial t'}+\vec{v}^{\;\prime}\cdot\nabla'\right) v_\Vert^{\prime\prime} +
\frac{e}{m_b} \; \vec{e}_\Vert\cdot\left(\vec{E}^{\,\prime} + \frac{1}{e n_b} \; \nabla' p_\Vert - \frac{p_\Vert - p_\bot}{e n_b B} \; \nabla' B\right) +
\frac{1}{m_b n_b} \frac{\partial\pi_{\Vert,\beta}}{\partial x'_\beta} =
\vec{v}^{\;\prime\prime}_\bot\cdot\left(\frac{\partial}{\partial t} + \vec{v}^{\;\prime}\cdot\nabla'\right)\vec{b},
\end{equation}
In the above, $\gamma_u$ is the relativistic factor of the unperturbed beam motion, $\gamma_u = 1/\sqrt{1-u^2/c^2}$ and $m_b$ is the relativistically increased {'perpendicular mass'} of the beam electrons, $m_b = \gamma_u m_0$. The stress of the beam electrons ${\bm\pi}_b$ and their pressure ${\bf p}_b$ are off-diagonal and diagonal tensors, respectively. The latter is given by ${\bf p} = p_\bot({\bf I} - \vec{e}_\Vert\vec{e}_\Vert) + p_\Vert \;\vec{e}_\Vert\vec{e}_\Vert$, where ${\bf I}$ is a unit tensor, viz. $I_{\alpha, \beta} = \delta_{\alpha, \beta}$ and $\delta_{\alpha, \beta}$ is the Kronecker delta, the directions parallel and perpendicular to the magnetic field are denoted by the subscripts $\Vert$ and $\bot$, respectively, and the standard vector algebra shorthand is used for the divergence of a tensor ${\bf w}$, viz. $\nabla\cdot{\bf w} = \vec{e}_\alpha(\partial w_{\alpha, \beta}/\partial x_\beta)$, and also for $\vec{p}\cdot\vec{q} \, \vec{r} = (\vec{p}\cdot\vec{q})\vec{r}$ and $\nabla\cdot\vec{q} \, \vec{r} = (\nabla\cdot\vec{q} + \vec{q}\cdot\nabla)\vec{r}$. A pressure anisotropy implies different parallel and perpendicular beam temperatures. The ratio of the unperturbed values of the pressures, observed in the laboratory frame, is given by $p_{\Vert_0}/p_{\bot_0} = \gamma_u^2 \; T'_{b \Vert_0}/T'_{b\bot_0}$, where the unperturbed temperatures $T'_{b \Vert_0}$ and $T'_{b\bot_0}$ are measured in the moving frame, see e.g. Ref. \cite{unmagnetized_beam}.%2013PhPl...20d4501N,reiser,

In a magnetized plasma, the closure of fluid equations (\ref{beam_cont}) and (\ref{beam_mom}) and the formulation of appropriate equations for $p_\Vert$, $p_\bot$, and $\pi_{\alpha,\beta}$ is relatively simple under the drift scaling and in the regime of small perturbations %of the density and
of the magnetic field, weak dependence along the magnetic field, and small but finite Larmor radius corrections, viz.
\begin{equation}\label{drift_scaling}
\varepsilon  \sim
\frac{u\;\partial/\partial z'}{\Omega_p} \sim
\frac{\partial/\partial t}{\Omega_p} \sim
\frac{\partial/\partial t'}{\Omega_b} \sim
\frac{\vec{v}^{\;\prime}\cdot\nabla'}{\Omega_b} \sim  \frac{|\vec{B}-\vec{e}_z B_0|}{B_0} \sim \frac{\vec{e}_\Vert\cdot\nabla'}{\left|\nabla_\bot'\right|}
{\lesssim}
\frac{p_\bot}{m_b n_b}\frac{{\nabla'_\bot}^{\!\! 2}}{\Omega_b^2} \sim \gamma_u^{-2} \ll 1,
\end{equation}
where $\Omega_p$ and $\Omega_b$ are the cyclotron frequencies of the (light) plasma and (heavy) beam electrons, respectively, viz. $\Omega_{p,b} = -e B_0/m_{p,b}$. We further assume that parallel heat fluxes
%$q_{\Vert,\Vert}$  and $q_{\bot,\Vert}$
are much smaller than the perpendicular fluxes, %$\vec{q}_{\Vert,\bot}$  and $\vec{q}_{\bot,\bot}$,
which is the applicability condition for the standard Grad's 13 moment approximation \cite{0bb4d204b6ac496a9782dc4b155cc02b} and is intuitively expected to hold for long beams, $L_\Vert\gg L_\bot$. Within such scaling we may neglect the gyroviscous part of the pressure tensor, see Eq. (4.136) in \cite{Fitzpatrick} and the fourth-order moments of the Vlasov equation. As the pressure is invariant under Lorentz transformations \cite{1971AmJPh..39..938C}, equations for parallel and perpendicular pressure components [see e.g. the nonrelativistic equations (60) and (64) in Ref. \cite{2008PhPl...15h2106R}], with the accuracy to the leading order in $\varepsilon\ll 1$ defined in Eq. (\ref{drift_scaling}), are simplified to
\begin{eqnarray}
&&
\frac{\partial p_\bot}{\partial t'} + \nabla'\cdot\left(p_\bot \vec{v}^{\;\prime}\right) + p_\bot\left[\nabla'\cdot\vec{v}^{\;\prime}-\vec{e}_\Vert\cdot\left(\vec{e}_\Vert\cdot\nabla'\right) \vec{v}^{\;\prime} \right] +
\nabla'\cdot \vec{q}_\bot
%{+ 2 \vec{q}_\Vert\cdot\left(\vec{e}_\Vert\cdot\nabla'\right) \vec{e}_\Vert}
= 0,
\label{perp_press}
\\
&&
\frac{1}{2}\left[\frac{\partial p_\Vert}{\partial t'} + \nabla'\cdot\left(p_\Vert \vec{v}^{\;\prime}\right)\right] + p_\Vert\vec{e}_\Vert\cdot\left(\vec{e}_\Vert\cdot\nabla'\right) \vec{v}^{\;\prime} + \nabla'\cdot\vec{q}_\Vert
%{-2 \vec{q}_\Vert\cdot\left(\vec{e}_\Vert\cdot\nabla'\right) \vec{e}_\Vert}
= 0.
\label{paral_press}
\end{eqnarray}
%{Terms in blue color are probably of higher order, if $\vec{q}_\Vert$ is of order ${\cal O}(\varepsilon)$.}
The perpendicular fluxes of the perpendicular and parallel heats are small quantities of order $\lesssim{\cal O}(\varepsilon)$ and given by \cite{2008PhPl...15h2106R}
\begin{equation}\label{heat_fluxes}
\vec{q}_\bot = \frac{2p_\bot\,\vec{e}_\Vert}{m_b\Omega_b}\times\nabla'
\frac{p_\bot}{n_b},
\quad\quad
\vec{q}_\Vert = \frac{p_\bot\, \vec{e}_\Vert}{2 m_b\Omega_b}\times\nabla' \frac{p_\Vert}{n_b}.
\end{equation}
Finally, noting that the pressure is invariant under Lorentz transformations \cite{1971AmJPh..39..938C}, for the components %$\pi_{\alpha,\beta}$
of the collisionless stress tensor $\bm{\pi}_b$ we use the standard Braginskii's expressions \cite{1965RvPP....1..205B}
\begin{eqnarray}
\nonumber && \pi_{\beta,\beta} = - \pi_{\alpha,\alpha} = \left(p_\bot/2\Omega_b\right)\left(\partial v'_\alpha/\partial x_\beta' + \partial v'_\beta/\partial x_\alpha'\right), \\
\nonumber && \pi_{\alpha,\beta} = \pi_{\beta,\alpha} = \left(p_\bot/2\Omega_b\right)\left(\partial v'_\alpha/\partial x_\alpha' - \partial v'_\beta/\partial x_\beta'\right), \\
\nonumber && \pi_{\alpha,\Vert} = \pi_{\Vert,\alpha} = - \left(p_\bot/\Omega_b\right)\left(\partial v'_\beta/\partial x'_\Vert + \partial v''_\Vert/\partial x_\beta'\right), \\
\nonumber && \pi_{\beta,\Vert} = \pi_{\Vert,\beta} = \left(p_\bot/\Omega_b\right)\left(\partial v'_\alpha/\partial x'_\Vert + \partial v''_\Vert/\partial x_\alpha'\right), \\
&& \pi_{\Vert,\Vert} = 0,
\label{braginskii}
\end{eqnarray}
where we used the notation $\partial/\partial x'_\Vert = \vec{e}_\Vert\cdot\nabla'$, while $\alpha$ and $\beta$ are two mutually perpendicular directions that are perpendicular also to the magnetic field's direction $\vec{e}_\Vert$.
The perpendicular (to $\vec B$) fluid velocity $\vec{v}_\bot^{\;\prime}$ of the (heavy) beam electrons is now obtained when we multiply Eq. (\ref{beam_mom}) with $\vec{e}_\Vert\times$ and can be expressed as the sum of the $\vec{E}\times\vec{B}$, diamagnetic, anisotropic-temperature, stress-related,
% [sometimes also called the FLR (finite-Larmor-radius) drift],
and polarization drifts, viz.
\begin{equation}\label{drift_velocity}
\vec{v}_\bot^{\;\prime} = \vec{v}_E^{\;\prime} + \vec{v}_D^{\;\prime} + \vec{v}_A^{\;\prime} +
\vec{v}_\pi^{\;\prime} + \vec{v}_p^{\;\prime},
\end{equation}
where
\begin{eqnarray}
\label{drifts_big}
&&
\vec{v}_E^{\;\prime} = -\frac{\vec{e}_\Vert}{B}\times\vec{E}^{\;\prime}, \hspace{5mm}
\vec{v}_D^{\;\prime} = \frac{-\vec{e}_\Vert}{e n_b B}\times\nabla' p_\bot, \hspace{5mm}
\vec{v}_B^{\;\prime} = \frac{-p_\bot \vec{e}_\Vert}{e n B^2}\times\nabla' B,
\\
&&
\vec{v}_A^{\;\prime} = \frac{-\left(p_\Vert - p_\bot\right)\vec{e}_\Vert}{e n_b B}\times\left(\vec{e}_\Vert\cdot\nabla\right)\vec{e}_\Vert,\hspace{5mm}
%\vec{v}_A^{\;\prime} = \frac{p_\bot-p_\Vert}{e n_b B} \; \nabla'\times\vec{e}_\Vert,\hspace{5mm}
\vec{v}_\pi^{\;\prime} = \frac{-\vec{e}_\Vert}{e n_b B}\times\vec{e}_\alpha\frac{\partial\pi_{\alpha, \beta}}{\partial x_\beta}, \hspace{5mm}
\vec{v}_p^{\;\prime} = \frac{\vec{e}_\Vert}{\Omega_b}\times\left(\frac{\partial}{\partial t'} + \vec{v}^{\;\prime}\cdot\nabla'\right)\vec{v}^{\;\prime\prime}.
\label{drifts_small}
\end{eqnarray}
For completeness, in the above list we have included also the grad-$B$ drift velocity $\vec{v}_B^{\;\prime}$, although it does not appear explicitly in the expression (\ref{drift_velocity}), but it will emerge later in the continuity equation by virtue of the term $\nabla'\cdot\vec{v}_\bot^{\;\prime}$.

Within the adopted scalings, Eq. (\ref{drift_scaling}), the electromagnetic field can be expressed in terms of the effective potential $\varphi\equiv\phi-uA_z$ and of the $z$-components of the vector potential and magnetic field, $A_z$ and $\delta B_z$, viz.
\begin{equation}
\label{E_B_field}\vec{E}' = -\nabla'\varphi - \frac{\partial}{\partial t'}\left(\vec{e}_z A_z + \vec{e}_z\times\nabla'_\bot \; {\nabla'_\bot}^{\!\!\! -2} \delta B_z\right),
\quad\quad
\vec{B} = \vec{e}_z \left(B_0 + \delta B_z\right) - \vec{e}_z\times\nabla'_\bot A_z.
\end{equation}
Noting from the ordering assumed in Eq. (\ref{res_lin_eq}) that $|\vec{\delta B}_\bot|/B_0\sim n_b/n_{p0}\sim (c/\vec{v}_\bot)(\delta B_z/B_0)\ll 1$, from the above  we obtain the following leading-order expressions
\begin{equation}\label{pppp}
\vec{e}_\Vert = \vec{e}_z - \frac{1}{B_0} \; \vec{e}_z\times\nabla'_\bot A_z,
\quad
B = |\vec{B}| = B_0, % + \delta B_z,
\quad
\vec{e}_\Vert\cdot\vec{E} = -\vec{e}_\Vert\cdot\nabla'\varphi - \frac{\partial A_z}{\partial t'},
\quad
\vec{e}_\Vert\times\vec{E} = -\vec{e}_z\times\nabla'_\bot\varphi,
\end{equation}
{
Using Eqs. (\ref{braginskii}) the stress' contribution to the parallel force in Eq. (\ref{beam_mom}) and the stress-related drift take the form:
\begin{eqnarray}
&&
\hspace{-.3cm}
\frac{\partial\pi_{\Vert,\beta}}{\partial x'_\beta} =
\vec{e}_\Vert\cdot\left[\nabla'_\bot v''_\Vert\times\nabla'_\bot\frac{p_\bot}{\Omega_b} - \nabla'_\bot\times\left(\frac{p_\bot}{\Omega_b} \frac{\partial\vec{v}_\bot^{\;\prime}}{\partial x'_\Vert}\right)\right].
\label{contr_stress}
\\
\nonumber
&&
\hspace{-.3cm}
\vec{v}_\pi^{\;\prime}  = \frac{-1}{e n_b B}\left\{\frac{\partial}{\partial x'_\alpha}\left(\frac{p_\bot}{2\Omega_b}\; \frac{\partial \vec{v}_\bot^{\;\prime}}{\partial x'_\alpha} \right)
+
\left[\left(\vec{e}_\Vert\times\nabla'_\bot\frac{p_\bot}{2\Omega_b} \right) \cdot \nabla'_\bot\right]\vec{e}_\Vert\times\vec{v}_\bot^{\;\prime} +
\frac{\partial}{\partial x'_\Vert}
\left[\frac{p_\bot}{\Omega_b} \left(\frac{\partial \vec{v}_\bot^{\;\prime}}{\partial x'_\Vert} + \nabla'_\bot v''_\Vert \right) \right] +
\frac{\partial\vec{v}_\bot^{\;\prime}}{\partial x'_\Vert} \frac{\partial}{\partial x'_\Vert}\frac{p_\bot}{2\Omega_b}\right\},
\\
\label{U_pi}
\end{eqnarray}
where $v''_\Vert = \vec{e}_\Vert\cdot \vec{v}^{\;\prime\prime}$ and we made use of $\vec{v}_\bot^{\;\prime\prime}\approx \vec{v}_\bot^{\;\prime}$. For an incompressible flow that is homogeneous along the magnetic field, the stress-related drift simplifies to
\begin{equation}\label{U_pi_simp}
\vec{v}_\pi^{\;\prime}  = \vec{e}_\Vert\times\left\{\frac{1}{e n_b B} \; \nabla'_\bot \left[\frac{p_\bot}{2\Omega_b}\;\nabla'_\bot\cdot \left(\vec{e}_\Vert \times\vec{v}_\bot^{\;\prime}\right)\right] - \frac{1}{\Omega_b} \left(\vec{v}_D^{\;\prime}\cdot\nabla'_\bot\right)\vec{v}_\bot^{\;\prime} \right\}.
\end{equation}
Using the above and the expressions for the perpendicular heat fluxes of perpendicular an parallel heats, Eqs. (\ref{heat_fluxes}), we rewrite the equations of continuity and parallel momentum, (\ref{beam_cont}) and (\ref{parallel_momentum_eq}), as well as the energy equations (\ref{perp_press}) and (\ref{paral_press}) in the following convenient form
\begin{eqnarray}
&&
\label{cont_eq_1}
\left(\frac{\partial}{\partial t'} + \vec{v}^{\;\prime}\cdot\nabla'\right) \log\frac{n_b}{B} +
\frac{\partial v'_\Vert}{\partial x'_\Vert} + \frac{1}{\Omega_b}\;\nabla'\cdot\left(\vec{e}_\Vert\times
\left\{\left[\frac{\partial}{\partial t'} +
\left(\vec{v}^{\;\prime}-\vec{v}_D^{\;\prime} + \vec{v}_B^{\;\prime}\right) \cdot \nabla'\right]\vec{v}^{\;\prime}\right\}\right) = 0,
\\
&&
\label{parallel_momentum_eq_1}
\left[\frac{\partial}{\partial t'} + \left(\vec{v}^{\;\prime}-\vec{v}_D^{\;\prime} + \vec{v}_B^{\;\prime}\right)\cdot\nabla'\right] v_\Vert^{\prime\prime} =
\frac{e}{m_b}\frac{\partial\varphi}{\partial x'_\Vert} - \frac{1}{e n_b}\left[\frac{\partial p_\Vert}{\partial x'_\Vert} + \nabla'\cdot\left(\frac{p_\bot}{\Omega_b}\; \vec{e}_\Vert\times \frac{\vec{v}_\bot^{\;\prime}}{\partial x'_\Vert}\right)\right].
\\
&&
\label{perp_press_1}
\left[\frac{\partial}{\partial t'} + \left(\vec{v}^{\;\prime}-\vec{v}_D^{\;\prime} + \vec{v}_B^{\;\prime}\right) \cdot\nabla'\right] \log\frac{p_\bot}{n_b^2}-\frac{\partial v'_\Vert}{\partial x'_\Vert} + \vec{v}_B^{\;\prime} \cdot\nabla'\log{p_\bot} = 0,
\\
&&
\label{paral_press_1}
\left[\frac{\partial}{\partial t'} + \left(\vec{v}^{\;\prime}-\vec{v}_D^{\;\prime} + \vec{v}_B^{\;\prime}\right) \cdot\nabla'\right] \log\frac{p_\Vert}{n_b}+ 2\, \frac{\partial v'_\Vert}{\partial x'_\Vert} = 0.
\end{eqnarray}
}
Within the drift ordering with weak parallel dependence and for perturbations that are much bigger than the Larmor radius, Eq. (\ref{drift_scaling}), the $\vec{E}\times\vec{B}$ and diamagnetic drifts are bigger than the anisotropic-temperature, polarization, and stress-related drifts, except the first term of the latter which is of the same order, see Eq. (\ref{U_pi}). However, our problem involves large variations of the beam density $n_b$, which is small but finite close to the beam axis, see Eqs. (\ref{Fourier_wave_eq}) and (\ref{res_lin_eq}), and vanishes at large distances. To accommodate such large relative density variations and and still have the diamagnetic drift of the same order as the $\vec{E}\times\vec{B}$ drift, we have to restrict our analysis to relatively cold beames, viz.
\begin{equation}\label{beam_T_scaling}
p_\bot/n_b \lesssim e\varphi \sim \varepsilon \, m_0c^2.
\end{equation}
Under such more restrictive scaling, our equations are further simplified since FLR corrections and the convection by the anisotropic temperature- and grad-$B$ drifts (i.e. terms $\vec{v}_A^{\;\prime}\cdot\nabla'$ and $\vec{v}_B^{\;\prime} \cdot \nabla'$) can be neglected, and also $\log B/B_0\ll\log n_b$. We further note that the condition $\phi \sim u A_z$ and, consequently,
\[
u (\vec{e}_\Vert-\vec{e}_z)\cdot\nabla'_\bot \lesssim \vec{v}_\bot^{\;\prime}\cdot\nabla'_\bot
\sim \partial/\partial t' \ll \partial/\partial t \sim u\;\partial/\partial z',
\]
implies $\vec{e}_\Vert = \vec{e}_z + {\cal O}(\varepsilon^2)$, that further implies $\vec{e}_\Vert\cdot\nabla' \to \partial/\partial z'$, $\vec{v}_\bot^{\;\prime\prime}\to\vec{v}_\bot^{\;\prime}$, and $v_\Vert^{\prime\prime}\to \gamma_u^2v_\Vert^{\prime}$, see Eq. (\ref{beam_mom}). It is also worth noting that in Eq. (\ref{cont_eq_1}) we may safely set $\log B\to 0$, since
\begin{equation}\label{B_scaling}
\log \delta B_z\sim\delta B_z/B_0\sim (v'_\bot/c)(n_b/n_{p0})\lesssim\varepsilon^2.
\end{equation}
Now, the equations of continuity, parallel momentum, and of the parallel and perpendicular energies (appropriately combined with the continuity equation) can be written in a simple form
\begin{eqnarray}
&&
\label{mom_eq_2}
%\left[{\cal D}_{t'} + \left(\vec{v}_\pi^{\;\prime} + \vec{v}_p^{\;\prime}\right)\cdot\nabla'_\bot\right]
{\cal D}_{t'}\log n_b + \frac{\partial v'_\Vert}{\partial z'}  - \frac{1}{B\Omega_b}\;\nabla'_\bot\cdot\left[{\cal D}_{t'}\left(\nabla'_\bot\varphi -\frac{\nabla'_\bot p_\bot}{e n_b}\right)\right] = 0,
\\
&&
\label{paral_mom_2}
{\cal D}_{t'} \, v'_\Vert =
\frac{e}{\gamma_u^2 \,m_b}\left(\frac{\partial\varphi}{\partial z'} - \frac{1}{e n_b} \frac{\partial p_\Vert}{\partial z'}\right)
\\
&&
\label{perp_press_2}
{\cal D}_{t'} \log\frac{p_\bot}{n_b^2} - \frac{\partial v'_\Vert}{\partial z'} = 0,
\\
&&
\label{paral_press_2}
D_{t'}\log\frac{p_\Vert}{n_b}  + 2 \; \frac{\partial v'_\Vert}{\partial z'} = 0,
\quad\quad
{\rm where}
\quad\quad
{\cal D}_{t'}  = \frac{\partial}{\partial t'} + \frac{1}{B_0}\left(\vec{e}_z\times\nabla' \varphi\right)\cdot\nabla' .
\end{eqnarray}
Note that we have $\log n_b = {\cal O}(\varepsilon^0)$, but the ordering of Eq. (\ref{drift_scaling}) still permits us to drop the polarization contributions to the convective derivative of $\log n_b$ and to keep the small but finite FLR corrections in the last term of Eq. (\ref{mom_eq_2}).

The equations for the perpendicular and parallel pressures (\ref{perp_press_2}) and (\ref{paral_press_2}) are coupled via the term $\partial v'_\Vert/\partial x'_\Vert$. In the isotropic case, $p_\Vert = p_\bot\equiv p$ their linear combination yields the standard equation of state for a 3-D adiabatic process
\begin{equation}
\label{press_3}
D_{t'}\log\frac{p}{n_b^{\kappa}} = 0,
\quad\Rightarrow\quad
p = \tau \, n_b^{\kappa},
\quad {\rm where}\quad
\kappa\equiv\kappa_\Vert=\kappa_\bot = {5}/{3},
\quad {\rm and}\quad
\tau\equiv\tau_\Vert =\tau_\bot = {\rm constant}.
\end{equation}
Conversely, noting from Eqs. (\ref{paral_mom_2}) and (\ref{drift_scaling}) that coupling between equations for $p_\bot$ and $p_\Vert$ is weak, since we have
\begin{equation}\label{coupling_ener_eq}
%{\cal D}_{t'}
\log\frac{p_\bot}{n_b^2} \sim
%D_{t'}
\log\frac{p_\Vert}{n_b} \sim
%D_{t'}
\frac{e}{m_b c^2}\left[\varphi - \left(\frac{\partial}{\partial z'}\right)^{-1}\left(\frac{1}{e n_b} \frac{\partial p_\Vert}{\partial z'}\right)\right] \sim
%{\cal D}_{t'}\;
{\cal O}\left(\varepsilon\right),
\end{equation}
for a sufficiently long beam we can write the following leading solutions, rigorously valid only if $\partial v'_\Vert/\partial z\to 0$, viz.
\begin{equation}\label{adiabatic_vert_bot}
p_\bot = \tau_\bot \, n_b^{\kappa_\bot},
\quad
p_\Vert = \tau_\Vert \, n_b^{\kappa_\Vert},
\quad {\rm where}\quad
\kappa_\bot= 2,
\quad
\kappa_\Vert= 1,
\quad {\rm and}\quad
\tau_\bot, \tau_\Vert = {\rm constant}.
\end{equation}

\section{Stationary states and their stability}
One easily verifies that our model equations (\ref{mom_eq_2})-(\ref{paral_press_2}) are satisfied by arbitrary stationary and $z$-independent functions $\phi, A_z, n_b, v'_\Vert, p_\Vert$, and $p_\bot$ that are cylindrically symmetric,
\begin{equation}\label{trivial}
\partial/\partial t'= \partial/\partial z'= \partial/\partial \theta'= 0,
\end{equation}
where $\theta'=\arctan\,(y'/x')$. Then, Eq. (\ref{plasma_equation}) yields an infinite number of solutions for the potentials, depending on our choice of the relation $n_b = n_b(\varphi)$. The number of solutions that are stable to small, but finite, perturbations is likely to be much smaller. To find possible stable solutions, we seek stationary states that exist when some of the conditions (\ref{trivial}) are not satisfied, i.e. if a certain class of perturbations is applied to a cylindrically symmetric, $z$-independent stationary structure. This can be done in two important regimes, described below, in which relatively simple analytic, albeit fully nonlinear solutions for the beam density $n_b$ can be obtained from Eqs. (\ref{mom_eq_2})-(\ref{paral_press_2}).

\subsection{Thermalized beam that is inhomogeneous in the direction of propagation}
\begin{figure}[htb]
\centering
\includegraphics[width=70mm]{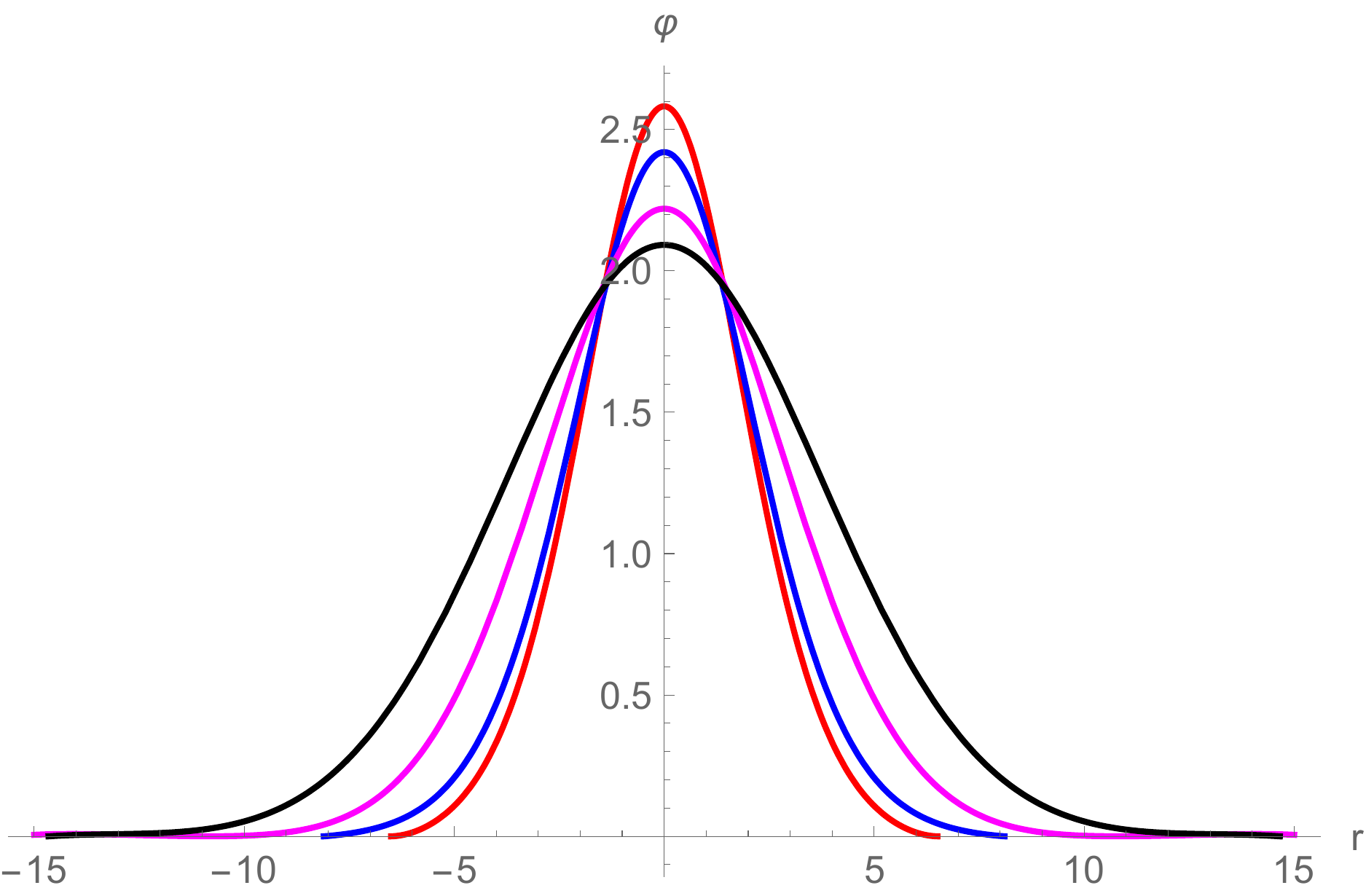}%{thermalized.pdf}%
\caption{Electrostatic potential $\varphi$ %and the beam density $n_b$ ($n_b=\varphi^\frac{3}{2}$)
of cylindrically symmetric stationary structures, found as the solutions of Eq. (\ref{plasma_equation_Boltzmann_dimensionless}). The red, blue, magenta, and black lines correspond to the solutions with different guide magnetic fields, for which the parameter $\Omega_p^2$ has the values $\Omega_p^2 = 0.1,\; 0.5, \; 3,\; {\rm and} \; 10$, respectively. Normalizations are given in Eq. (\ref{normalization_Boltzmann}).} \label{potential_Boltz}
\end{figure}

We seek a solution whose effective parallel phase velocity is sufficiently small so that we may neglect the left-hand side of the parallel momentum equation (\ref{paral_mom_2}). {Estimating $v'_\Vert$ from Eq. (\ref{mom_eq_2}), we find that this can be done when
\begin{equation}\label{Therm_beam}
\left(\frac{{\cal D}_{t'}}{\partial/\partial z'}\right)^2\ll\frac{p_\Vert}{\gamma_u^2 \, m_b n_b}
\end{equation}
For such structure, that is fittingly inhomogeneous in the direction of propagation, we also have a fully 3-D adiabatic process, which is in the case isotropic pressure described by Eq. (\ref{press_3}).} Then, Eq. (\ref{paral_mom_2}) yields the beam density as
\begin{equation}\label{Boltzmannean_beam}
n_b =\left( -\frac{2}{5}\;\frac{e\varphi}{\tau}\right)^\frac{3}{2}.
\end{equation}
Introducing the following normalized quantities
\begin{equation}\label{normalization_Boltzmann}
\varphi\to -e\varphi \left(\frac{2}{5 \tau}\right)^3 \left(\frac{m_0 c^2}{n_{p0}}\right)^2, \quad
r\to \frac{\omega_{pp} r'}{c}, \quad
\Omega_p\to\frac{\Omega_p}{\omega_{p p}},
\end{equation}
our equations (\ref{plasma_equation}) and (\ref{Boltzmannean_beam}) simplify to the following dimensionless equation
\begin{equation}\label{plasma_equation_Boltzmann_dimensionless}
\left[\left(1-\nabla_\bot^2\right)^2+\Omega_p^2\nabla_\bot^4\right]  \varphi + \left(1-\nabla_\bot^2\right) \varphi^\frac{3}{2} = 0,
\end{equation}
that is solved numerically in a cylindrically symmetric case, $\varphi = \varphi(r)$ and the solution is displayed in Fig. \ref{potential_Boltz}. Apart from certain radial spreading, the solution %is localized by the nonlinearity $\propto\varphi^{3/2}$ and
is similar to that obtained in the absence of the guide magnetic field  \cite{unmagnetized_beam}.

\subsection{Helicoidally shaped monoenergetic beam}
In the case of a beam with a very narrow velocity spread in the parallel direction, i.e. in the regime with $\tau_\Vert\to 0$ and $\tau_\bot\ne 0$, we have a 2-D adiabatic process in the transverse direction with $\kappa_\bot = 2$. For such monoenergetic beam we seek a corkscrew shaped (i.e. helicoidal) solution that is expected to emerge when the beam is launched at a small pitch angle to the magnetic field, so that the Lorentz force wraps it around the $z$ axes in the form of a helix. More specifically, we seek a solution that is propagating along the $z$ axis with the nonrelativistic phase velocity $\delta u$ and spinning around it with the angular velocity $\omega_b$, satisfying $\omega_b\ll\Omega_b$. Such solution depends only on the variables  $r''=\sqrt{{x'}^2+{y'}^2}$ and  $t'' = t'- z'/\delta u - \theta'/\omega_b$, where $\theta'=\arctan\,(y'/x')$. {The periodicity length of such helix in the $z$ direction is $L_z'=\delta u/\omega_b$, and the frequency of this structure observed in the laboratory frame is given by
\begin{equation}\label{frequency_lab}
\omega = (2\pi/L_z')\, u =2\pi (u/\delta_u)\,\omega_b \ll \Omega_p,
\end{equation}
which agrees with the plasma scaling (\ref{frequency_scaling}) when $u/\delta u>\gamma_u$. It should be noted also that the self-modulation instability in a magnetized plasma, described in our earlier papers \cite{2014EPJD...68..271F,2014EPJD...68..210F} arises through the nonlinear convection coming from the curlfree part of the fluid velocity, see e.g. Refs. \cite{Kaufman-Stenflo,1976PhFl...19..872P,1977PlPh...19..889Y}. Under the drift scaling Eq. (\ref{drift_scaling}), the latter velocity component is essentially equal to the polarization drift and the resulting nonlinear terms are small quantities of a higher order, not included in our basic equations (\ref{mom_eq_2})-(\ref{paral_press_2}). In other words, the self-modulation and the concomitant betatron emission occur on a time scale that is slow (in the comoving frame) compared to the rotational frequency of the helicoidal beam,  $\omega_b$, and to the related oscillations observed in the laboratory frame described by Eq.  (\ref{frequency_lab}).
}

Now, for a helicoidal moving/spinning structure the parallel momentum equation takes the form
\begin{equation}\label{traveling_par_mom}
\left[\vec{e}_z\times\nabla''\left(\varphi - {B_0\omega_b {r''}^2}\!\! /\,{2}\right)\right]\cdot\nabla''\left(v'_\Vert + \frac{e}{m_b} \frac{\varphi}{\gamma_u^2\,\delta u}\right) = 0,
\end{equation}
which is readily integrated as
\begin{equation}\label{res_par_mom}
v'_\Vert = -\frac{e}{m_b}\frac{\varphi}{\gamma_u^2 \, \delta u} + \delta u \; {\cal F}\left(\varphi - {B_0\omega_b{r''}^2}\!\!/\,{2}\right),
\end{equation}
where $\cal F$ is an arbitrary function of its argument. Substituting these in Eq.(\ref{mom_eq_2}) and using $p_\bot = \tau_\bot \, n_b^2$ we have
\begin{eqnarray}
\nonumber
&&
\left[\vec{e}_z\times\nabla''\left(\varphi - {B_0\omega_b{r''}^2}\!\!/\,{2} \right)\right]\cdot\nabla''
\left\{\log\,n_b +
\left[\frac{e}{m_b} \frac{1}{\gamma_u^2\, \delta u^2} - {\cal F}\,'\left(\varphi - {B_0\omega_b{r''}^2}\!\!/\,{2}\right) \right]\, \varphi \right\} +
 \\
&&
\frac{1}{\Omega_z B_0}
\nabla''\cdot \left(\left\{\left[\vec{e}_z\times\nabla'' \left(\varphi - {B_0\omega_b{r''}^2}\!\!/\,{2}\right) \right]\cdot\nabla''\right\}
\nabla''\left(\varphi-\frac{2\tau_\bot}{e}\; n_b\right)\right)
= 0,
\label{trav_cont_eq}
\end{eqnarray}
where ${\cal F}\,'(\xi)=d{\cal F}(\xi)/d\xi$.
We conveniently introduce the following dimensionless quantities
\begin{eqnarray}\nonumber
&&
\varphi\to \frac{e\varphi}{T_0}, \quad
n_b\to\frac{m_0 c^2}{T_0}\frac{n_b'}{n_{p 0}}, \quad
\vec{r}\to \frac{\omega_{pp} \vec{r}\,''}{c},
\\
&&
{\omega_b\to\frac{m_0 c^2}{T_0}\frac{\Omega_p\,\omega_b}{\omega_{pp}^2} ,}\quad
\Omega_p\to\frac{\Omega_p}{\omega_{p p}}, \quad
\tau_\bot\to \frac{\tau_\bot n_{p 0}}{m_0 c^2}, \quad
\rho = \frac{\omega_{p p}}{\Omega_p} \left({\frac{\gamma_u T_0}{m_0 c^2}}\right)^\frac{1}{2}, \quad
d =\frac{c}{\gamma_u^2 \delta u}\;\frac{\Omega_p}{\omega_{p p}},
\label{normalization}
\end{eqnarray}
where $T_0$ is a constant of normalization, adopted so that in the spatial region of interest we have $\varphi\sim n_b\sim\omega_b\sim{\cal O}(1)$. Now we can rewrite Eq. (\ref{trav_cont_eq}) in a normalized form as
\begin{eqnarray}
\nonumber
&&
\left[\vec{e}_z\times\nabla\left(\varphi - \omega_b r^2\!\!/\,2 \right)\right]\cdot\nabla
\left\{\log\,n_b + \rho^2 d^2
\varphi \, \left[1 - {\cal F}\,'\left(\varphi - \omega_b r^2 \!\!/\,2\right) \right]\right\} +
\\
&&
\rho^2 \nabla_\bot\cdot \left(\left\{\left[\vec{e}_z\times\nabla_\bot\left(\varphi - \omega_b r^2 \!\!/\,2\right) \right]\cdot\nabla_\bot\right\}
\nabla_\bot\left(\varphi-2\tau_\bot n_b\right)\right)
= 0,
\label{trav_cont_dimensionless}
\end{eqnarray}
In the rest of the paper we use ${\cal F}\,'(\xi) = 0$, which yields a simple (linear) expression for $v'_\Vert$, viz. $v'_\Vert =  - e \varphi/(m_b \gamma_u^2 \, \delta u)$. Such choice is equivalent to that we have used in the pressure equations, see Eqs. (\ref{press_3}) and (\ref{adiabatic_vert_bot}).

We note that for a nonrelativistic phase velocity $\delta u$ the parameter $d$ scales as $d = 2\, (\Omega_p / \omega_{pp})(c-u)/\delta u \sim {\cal O}(1)$, while the parameter $\rho$ is very small, viz. $\rho^2 = (v_{T0}/c)^\frac{3}{2} [v_{T_0}/2(c-u)]^\frac{1}{2} (\omega_{pp}^2/\Omega_p^2)\ll 1$, where $v_{T 0}$ is the thermal velocity of electrons with the temperature $T_0$, viz. $v_{T 0} = (T_0/m_0)^\frac{1}{2}$.

In the limit $\rho\to 0$ the above equation is solved for the beam density as $n_b\to {\cal G}(\varphi - \omega_b r^2\!\!/\,2)$ where ${\cal G}$ is an arbitrary function of its argument. Then, with the accuracy to the first order in the small parameter $\rho^2$, using the identity
\begin{equation}\label{identity}
2 \, \nabla_\bot\cdot\left\{\left[\left(\vec{e}_z\times\nabla_\bot f\right)\cdot\nabla_\bot\right]\nabla_\bot G\left(f\right)\right\} = \left(\vec{e}_z\times\nabla_\bot f\right)\cdot\nabla_\bot\nabla_\bot^2 G\left(f\right) - \left[\vec{e}_z\times\nabla_\bot\nabla_\bot^2 f\right]\cdot\nabla_\bot G\left(f\right),
\end{equation}
and after some algebra, the continuity equation (\ref{trav_cont_dimensionless}) can be rewritten as a complete mixed product
\begin{equation}
\left\{\vec{e}_z\times\nabla_\bot \left[\varphi - \frac{\omega_b r^2}{2} + \tau_\bot n_b \; \rho^2 \nabla_\bot^2\left(\frac{3}{2}\;\varphi -\tau_\bot n_b \right) \right]\right\} \cdot\nabla_\bot
\left[\log\, n_b - \rho^2\nabla_\bot^2 \left(\varphi-\tau_\bot\, n_b \right)+
\rho^2 d^2 \,\varphi\right] = 0,
\label{iterative_n}
\end{equation}
which is readily integrated in terms of a continuous, but otherwise arbitrary function ${\cal G}(\xi)$, viz.
\begin{eqnarray}
\label{resenje_n_b}
n_b
&& \hspace{-3mm}
= {\cal G}\left[\varphi - \frac{\omega_b r^2}{2} + \tau_\bot n_b\;\rho^2 \nabla_\bot^2 \left(\frac{3}{2}\; \varphi - \tau_\bot n_b\right)\right] \times \exp\left[\rho_\bot^2 \nabla_\bot^2 \left(\varphi-\tau_\bot n_b\right) -
\rho^2 d^2 \;\varphi\right],
\\
&& \hspace{-3mm}
\approx {\cal G}\left(\varphi - \frac{\omega_b r^2}{2}\right) \left[1+\rho^2\nabla_\bot^2 \left(\varphi-\tau_\bot n_b^{(0)}\right) - \rho^2 d^2\varphi\right] +
{\cal G}'\left(\varphi - \frac{\omega_b r^2}{2}\right)
\tau_\bot n_b^{(0)} \, \rho^2\nabla_\bot^2 \left(\frac{3}{2}\; \varphi - \tau_\bot n_b^{(0)}\right).
\label{resenje_n_b_simpl}
\end{eqnarray}
The above expression for the beam density $n_b$ is to be used in the dimensionless version of the plasma equation (\ref{plasma_equation}), that with the use of the normalizations Eq. (\ref{normalization}), takes the form
\begin{equation}\label{plasma_equation_dimensionless}
\left[\left(1-\nabla_\bot^2\right)^2 + \Omega_p^2\nabla_\bot^4\right]  \varphi = \left(1- \nabla_\bot^2\right) n_b.
\end{equation}
Following the standard procedure for the construction of vortex solutions, we adopt ${\cal G}(\xi)$ to be a part-by-part linear function, whose slope $G_1$ is allowed to jump at some value of its argument, viz. ${\cal G}(\xi)=(\xi-\xi_0)\, G_1$ where $G_1 = G_1^{in,out}$ if $\xi \lessgtr\xi_0$, respectively. The condition $\xi(r,\theta)=\xi_0$ determines a closed line in the $r,\theta$ plane, that is usually referred to as the edge of the vortex core.
As the beam density $n_b$ contains only a small nonlinear term $\sim\rho^2$, we will seek the solution of Eq. (\ref{plasma_equation_dimensionless}) in the form $\varphi = \varphi_0(r) + \rho^2\delta \varphi_0(r) + \varphi_1(r)\,\cos\theta$, where $\varphi_1 \ll \varphi_0$ is the amplitude of the first cylindrical harmonic and $\delta \varphi_0(r)$ is a small nonlinear correction to the zeroth cylindrical harmonic.
In other words, we seek a solution that is predominantly cylindrically symmetric, but (slightly) displaced from the center $r=0$ and wrapped around the $z$ axes in the form of a helix. The vortex core in such a case needs to be adopted as a slightly shifted circle whose radius is $r_0$, viz. $r = r_0 + \delta r \cos\theta$, where $\delta r\ll r_0$; its 3-D shape is shown in Fig. \ref{separatrix}.

\paragraph{Outside solution}
From the condition that the beam density vanishes at $r\to\infty$ we readily see that outside of the vortex core we must have $G_1^{out}=0$ and, consequently, $n_b^{out} = 0$, see Eq. (\ref{resenje_n_b}). The solution for the potential outside the vortex core, $\varphi^{out}$, is determined from
\begin{equation}\label{Eq_out}
\left(\nabla_\bot^2-\kappa^2\right)\left(\nabla_\bot^2 - {\kappa^*}^{\,2}\right) \left(\varphi_0^{out} + \varphi_1^{out}\,\cos\theta\right) = 0,
\quad {\rm where} \quad
\kappa = \left(1+ i\,\Omega_p\right)^{-\frac{1}{2}},
\end{equation}
that is readily solved as $\delta\varphi_0^{out}=0$ and
\begin{equation}
\varphi_0^{out} = \alpha_{0_R}\, {\rm Re} \, K_0 \left(\kappa r\right) + \alpha_{0_I}\, {\rm Im}\, K_0 \left(\kappa r\right),
\quad\quad
\varphi_1^{out} = \alpha_{1_R}\, {\rm Re} \,K_1 \left(\kappa r\right) + \alpha_{1_I}\, {\rm Im}\,K_1 \left(\kappa r\right),
\label{inside}
\end{equation}
where $\alpha_{j_R}$ and $\alpha_{j_I}$ are arbitrary real constants, $K_j$ are modified Bessel functions, and $j=0,1$.

\paragraph{Inside solution}
Substituting the beam density, Eq. (\ref{resenje_n_b_simpl}), into the plasma equation (\ref{plasma_equation_dimensionless}) and expanding the latter in $\delta\varphi_1, \rho^2\ll 1$, we obtain the following equations %for $\varphi^{(0)}$, $\delta\varphi_0$, and $\delta\varphi_1$
\begin{eqnarray}
&&
\hspace{-.6cm}
\left(\nabla_\bot^2 + k_+^2\right)\left(\nabla_\bot^2 - k_-^2\right) \left(\varphi_0^{in} + a r^2 + b +\varphi_1^{in} \,\cos\theta\right) =0,
\nonumber%\label{plasma_eq_dim_hom}
\\
&&
\hspace{-.6cm}
\left(\nabla_\bot^2 + k_+^2\right)\left(\nabla_\bot^2 - k_-^2\right) \delta\varphi_0^{in} = n_{NL}^{in},
\label{plasma_eq_dim_inh}
\end{eqnarray}
where $G_1\to G_1^{in}$ and $n_{b0}^{in} = G_1\,(\varphi_0^{out} - \omega_b r^2/2-\xi_0)$ and
\[
n_{NL}^{in} = \frac{1- \nabla_\bot^2}{1+\Omega_p^2} \left\{n_{b0}^{in}\left[\nabla_\bot^2 \left(\varphi_0^{in} -\tau_\bot n_{b0}^{in}\right) - d^2\varphi_0^{in} +
G_1\, \tau_\bot \nabla_\bot^2 \left(3\,\varphi_0^{in}\!\!/2 - \tau_\bot n_{b0}^{in}\right)\right]\right\}
\]
\[
k_\pm^2 = \frac{1}{2\left(1+\Omega_p^2\right)}\left[\pm\left(G_1-2\right)+ \sqrt{G_1^2-4 \Omega_p^2\left(1-G_1\right)}\right],
\quad\quad
a=\frac{G_1\omega_b}{2\left(1-G_1\right)},
\quad\quad
b = \frac{G_1}{1-G_1}\left(\frac{2\omega_b}{1-G_1} + \xi_0\right).
\]
For later purposes, it is convenient to use $k_+$ as the constant of integration and to express $G_1$ and $k_-$ in terms of $k_+$
\[
G_1=1+k_+^2+ \frac{k_+^4\,\Omega_p^2}{1+k_+^2}
\quad\quad
k_-^2 = 1 - \frac{\Omega_p^2}{\left(1+k_+^2\right)\left(1+\Omega_p^2\right)}.
\]
We note that if we adopt $k_+^2>0$, we automatically obtain $k_-^2>0$. Equations %(\ref{plasma_eq_dim_hom}) and
(\ref{plasma_eq_dim_inh}) are readily solved
\begin{eqnarray}
\nonumber%\label{outside_0}
&&
\varphi_0^{in}=\beta_0\, J_0\left(k_+ r\right) + \gamma_0\, I_0\left(k_- r\right)-ar^2-b,
\\
\nonumber%\label{outside_1}
&&
\varphi_1^{in}=\beta_1\, J_1\left(k_+ r\right) + \gamma_1\, I_1\left(k_- r\right),
\\
&&
\label{nehomog_1}
%\hspace{-1cm}
\delta\varphi^{in}_0\left(r\right) = -K_0\left(k_- r\right)\int_{0}^r I_0\left(k_- r'\right) {f}_{NL}\left(r'\right) r' dr' -
I_0\left(k_- r\right)\int_r^{r_0} K_0\left(k_- r'\right) {f}_{NL}\left(r'\right) r' dr',
\end{eqnarray}
where
\begin{equation}
\label{nehomog_2}
%\hspace{-1cm}
f_{NL}\left(r\right) = \frac{\pi}{2}\left[J_k\left(k_+ r\right)
\int_r^{r_0} Y_k\left(k_+ r'\right) n_{NL}^{in}\left(r'\right) r' dr' +
Y_k\left(k_+ r\right)\int_0^r J_k\left(k_+ r'\right) n_{NL}^{in}\left(r'\right) r' dr'\right],
\end{equation}
$\beta_j$ and $\gamma_j$ with $j=0,1$ are arbitrary real constants, and $J_j, I_j$ are the Bessel functions and the modified Bessel functions of the first kind, respectively.

\begin{figure}[htb]
\centering
\includegraphics[width=70mm]{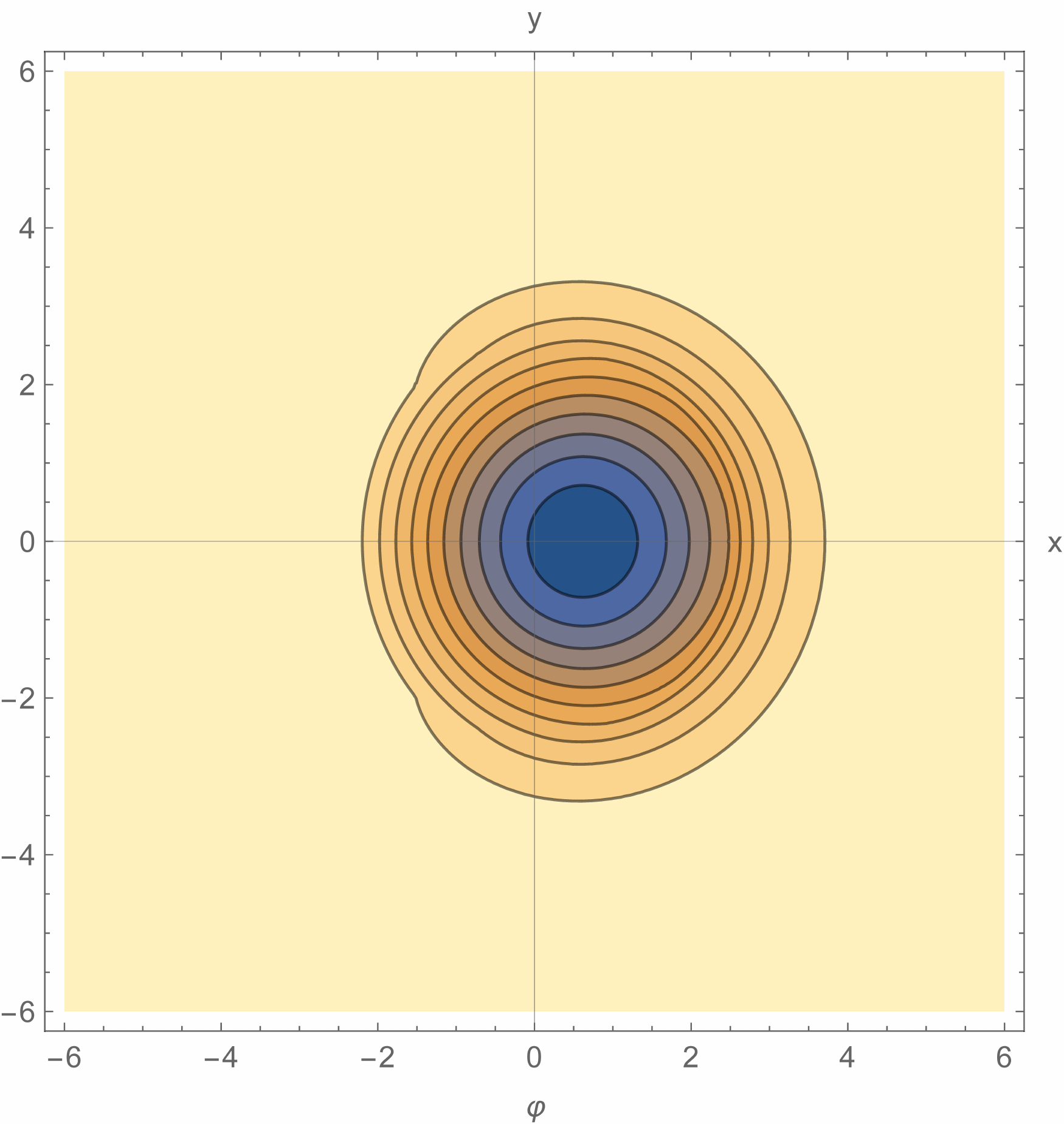}%{fusilli_phi_nb.pdf}%
\caption{Contour plot of the electrostatic potential $\varphi$
%(left) and of the beam density $n_b$ (right)
of a corkscrew-shaped, stationary rotating structure described by Eqs. (\ref{inside}) and (\ref{nehomog_1}), that rotates around the point $(x=0,y=0)$ and propagates along the $z$ axis. The normalizations are given in Eq. (\ref{normalization}) and the parameters of the solution and of its environment are adopted as
$\omega_b=2.5$,  $r_0=2.5$,  $\delta r=0.8$,  $\Omega_p=-0.7$.} \label{potential_helix}
\end{figure}

\begin{figure}[htb]
\centering
\includegraphics[width=100mm]{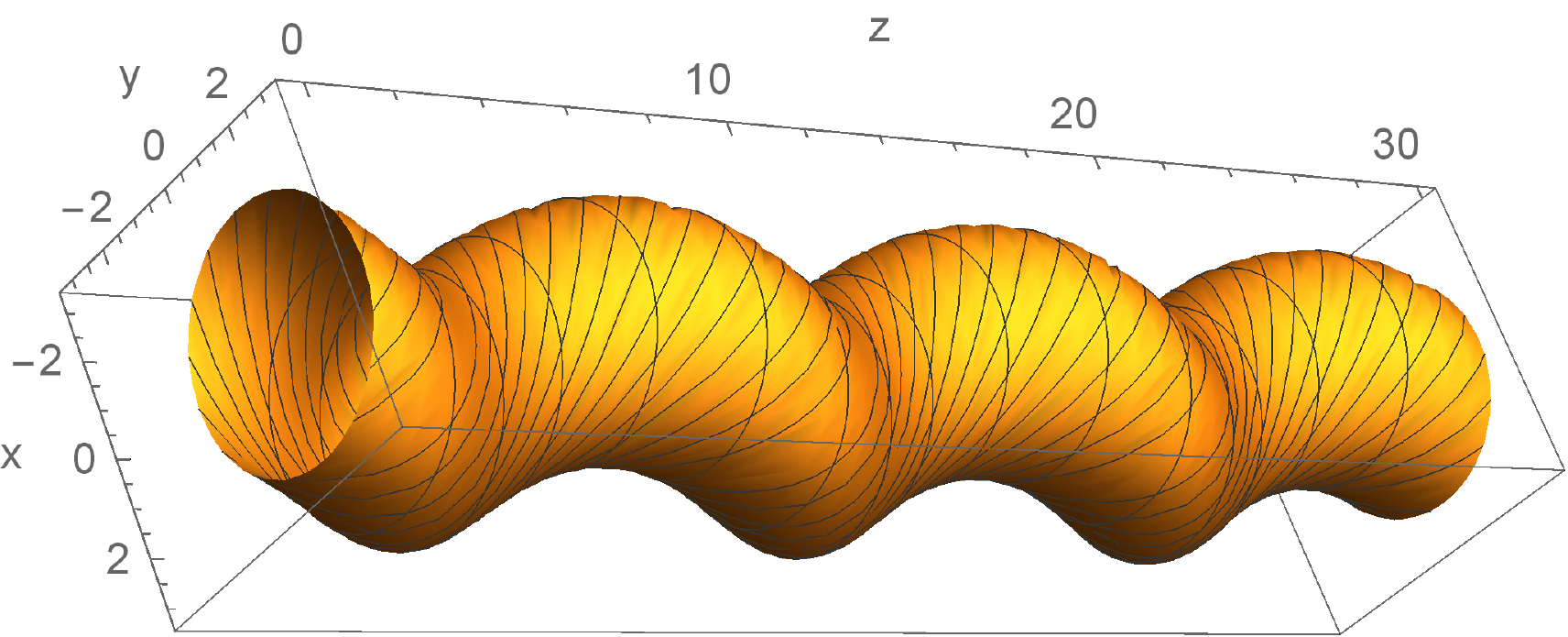}
\caption{Three-dimensional plot of the separatrix pertaining to the solution displayed in Fig. \ref{potential_helix}. The parallel phase velocity is adopted as $\delta u = \pi/2$.} \label{separatrix}
\end{figure}

The above "$in$" and "$out$" solution must be properly connected at the edge of the core. Separating the zeroth and the first cylindrical harmonics and with the accuracy to the first order in the small parameters $\delta r$ and $\rho^2$, the continuity of the function $\cal G$ and of the radial electric field gives the following six equations
\begin{eqnarray}
\label{cont_G_0}
&&
\varphi_0^{out}\left(r_0\right) = \varphi_0^{in}\left(r_0\right)+\delta\varphi_0^{in}\left(r_0\right) = \omega_b r_0^2/2 + \xi_0,
\\
&&
\label{cont_G_1}
\varphi_1^{out}\left(r_0\right) +\delta r\left(d/d r_0\right)\varphi_0^{out}\left(r_0\right) =
\varphi_1^{in}\left(r_0\right) +\delta r\left(d/d r_0\right)\varphi_0^{in}\left(r_0\right) = 0,
\\
&&
\label{cont_E_0}
\left(d/d r_0\right)\varphi_0^{out}\left(r_0\right) = \left(d/d r_0\right)\left[\varphi_0^{in}\left(r_0\right)+ \delta\varphi_0^{in}\left(r_0\right)\right],
\\
&&
\label{cont_E_1}
\left(d/d r_0\right)\varphi_1^{out}\left(r_0\right) + \delta r\left(d^2/d r_0^2 \right)\varphi_0^{out}\left(r_0\right) =
\left(d/d r_0\right)\varphi_1^{in}\left(r_0\right) +\delta r\left(d^2/d r_0^2\right)\varphi_0^{in}\left(r_0\right).
\end{eqnarray}
A close inspection reveals that the solutions ${\rm Im}\,K_j$ (outside) and $I_j$ (inside) must have smaller amplitudes than the other two  linearly independent modes. These solutions vanish in a sufficiently dense plasma, in the limit $\Omega_p\to 0$. Moreover, for a finite $\Omega_p$ the "$in$" and "$out$" solutions can be properly connected at the edge of the core only if the amplitude of the latter is sufficiently small, $\gamma_j<\beta_j$. For simplicity, we set $\alpha_{j_I}=\gamma_j=0$, and eliminate the remaining constants of integration from the continuity conditions (\ref{cont_G_0})-(\ref{cont_E_1}), which after some algebra yields
\begin{eqnarray}
\nonumber
&&
\left(G_1-1\right)\frac{d}{d r_0}\log\left[\frac{J_0\left(k_+ r_0\right)}{{\rm Re}\,K_0\left(\kappa r_0\right)}\frac{{\rm Re}\,K_0'\left(\kappa r_0\right)}{{\rm Re}\,K_1\left(\kappa r_0\right)}\right] =
\\
&&
\frac{d}{d r_0}\log\left[Q\left(r_0\right)\right]\times
\frac{\left(d/d r_0\right)\log\left[J_0'\left(k_+ r_0\right)/Q'\left(r_0\right)\right]}{\left(d/d r_0\right)\log\left[J_0\left(k_+ r_0\right)/Q\left(r_0\right)\right]}
\left\{\frac{\left(d/d r_0\right)\log\left[J_0\left(k_+ r_0\right)\right]}{\left(d/d r_0\right)\log\left[{\rm Re}\,K_0'\left(\kappa r_0\right)\right]}+G_1-1\right\},
\label{disp_relacija}
\end{eqnarray}
where $Q(r) = a_1(r^2-2 \xi_0/\omega_b) + b_1$ and the prime denotes the derivative of a function. As a typical example, we adopted the parameters of the solution and of its environment to be $\omega_b=2.5$,  $r_0=2.5$, $\delta r=0.8$, and $\Omega_p=-0.7$, and solved numerically the dispersion relation. The corresponding corkscrew-shaped, stationary rotating structure whose electrostatic potential is described by Eqs. (\ref{inside}) and (\ref{nehomog_1}) is displayed in Fig. \ref{potential_helix}.

The FLR terms appear only as corrections in the nonlinear dispersion relation (\ref{disp_relacija}), and they only slightly modify the radial profile of the structure in Fig. \ref{potential_helix}. We note also that a purely 2-D stationary state, with $\partial/\partial z = \partial/\partial\theta = 0$, may have an arbitrary radial dependence, while the above described helicoidal structure has a unique transverse profile, determined from the dispersion relation (\ref{disp_relacija}) for each particular radial extend $r_0$ and the pitch angle $r_0 \omega_b/\delta u$. In a more realistic case, allowing for a small but finite contribution of the linear solutions ${\rm Im}\,K_j$ (outside) and $I_j$ (inside), the radial dependence of the helicoidal structure would acquire some degree of freedom.

\section{Concluding remarks}
In this paper we have developed a hydrodynamic theory of a long relativistic beam propagation in an overdense plasma, in the regime in which the dynamics in the comoving frame can be regarded as nonrelativistic. Two different regimes of the evolution have been considered with respect to the rate of parallel heat convection. When the parallel phase velocity is well below the thermal speed, the beam electrons are effectively thermalized and the beam behavior is similar to that observed earlier in the nonmagnetized case \cite{unmagnetized_beam} with an additional degree of widening that comes from the increased spatial dispersion of electrostatic perturbations in a magnetized plasma. A stationary solution for the beam profile is found, whose localization in the transverse direction is governed by the nonlinearity associated with the 3-D adiabatic expansion of the beam. In the case when the parallel phase velocity is sufficiently large to prevent the heat convection along the magnetic field, we have demonstrated that the stationary state may have an arbitrary, cylindrically symmetric profile. The stability properties of the latter are assessed by studying a beam that has been launched at a small pitch angle relative to the magnetic field. We have demonstrated that the resulting helicoidal structure features a specific transverse profile whose shape is unique and fully determined by the radial extend $r_0$ and the pitch angle $\omega_b r_0/\delta u$, which indicates that the other arbitrary shapes permitted in a purely parallel propagation become unstable when the beam is not perfectly aligned.

\begin{acknowledgements} This work was supported in part (D.J. and M.B.) by the MPNTR 171006 and NPRP 8-028-1-001 grants. D.J. acknowledges financial support from the INFN's fondo FAI and the hospitality of the Dipartimento di Fisica "Ettore Pancini", Universita di Napoli "Federico II", Italy.

\end{acknowledgements}

%\bibliography{Beam_evolution_2016}

\end{document}